\documentclass[journal]{IEEEtran}

\usepackage{graphicx}
\usepackage{dcolumn}
\usepackage{bm}
\usepackage[utf8]{inputenc}
\usepackage[T1]{fontenc}
\usepackage{mathptmx}
\usepackage{csquotes}
\usepackage{subcaption}
\usepackage{hyperref}
\usepackage{xcolor}
\hypersetup{linktoc=all}
\usepackage{booktabs}
\usepackage{array}
\usepackage{multirow}
\usepackage{siunitx}
\usepackage{xcolor}
\usepackage{listings}
\usepackage{epstopdf}
\usepackage{hyperref}
\hypersetup{colorlinks, citecolor=green, filecolor=black, linkcolor=blue, urlcolor=blue }
\usepackage{graphicx}
\usepackage{amsmath}
\newcommand{\norm}[1]{\left\lVert#1\right\rVert}
\usepackage[version=4]{mhchem}
\usepackage{siunitx}
\usepackage{longtable,tabularx}
\newcolumntype{C}[1]{>{\centering\let\newline\\\arraybackslash\hspace{0pt}}m{#1}}
\usepackage{amsmath,amssymb}

\usepackage[flushleft]{threeparttable}
\usepackage{array,booktabs,makecell}
\usepackage{tikz}
\usetikzlibrary{shapes,arrows}
\usepackage{newfloat}
\DeclareFloatingEnvironment[fileext=dia,within=section]{diagram}

\usepackage{geometry}
\definecolor{darkgreen}{rgb}{0,0.6,0}

\begin{document}
	
	\title{Identifying Entangled Physics Relationships through Sparse Matrix Decomposition to Inform Plasma Fusion Design}
	
	\author{M.~Giselle~Fern\'andez-Godino, Michael~J.~Grosskopf, Julia~B.~Nakhleh, \\Brandon~M.~Wilson, John~Kline, and~Gowri~Srinivasan%
		\thanks{All authors are with Los Alamos National Laboratory, Los Alamos,
			NM, 87544 USA (corresponding author email: \href{mailto:mariagisellefernandez@gmail.com}{mariagisellefernandez@gmail.com}).}}
	
	\markboth{IEEE Transactions on Plasma Science,~Vol.~X, No.~X,~XXX~XXXX}%
	{Nakhleh \MakeLowercase{\textit{et al.}}: Exploring Sensitivity of ICF Outputs to Design Parameters in Experiments Using Machine Learning}
	
	\maketitle
	\begin{abstract}	
		A sustainable burn platform through inertial confinement fusion (ICF) has been an ongoing challenge for over 50 years. Mitigating engineering limitations and improving the current design involves an understanding of the complex coupling of physical processes. While sophisticated simulations codes are used to model ICF implosions, these tools contain necessary numerical approximation but miss physical processes that limit predictive capability. Identification of relationships between controllable design inputs to ICF experiments and measurable outcomes (e.g. yield, shape) from performed experiments can help guide the future design of experiments and development of simulation codes, to potentially improve the accuracy of the computational models used to simulate ICF experiments. We use sparse matrix decomposition methods to identify clusters of a few related design variables. Sparse principal component analysis (SPCA) identifies groupings that are related to the physical origin of the variables (laser, hohlraum, and capsule). A variable importance analysis finds that in addition to variables highly correlated with neutron yield such as picket power and laser energy, variables that represent a dramatic change of the ICF design such as number of pulse steps are also very important. The obtained sparse components are then used to train a random forest (RF) surrogate for predicting total yield. The RF performance on the training and testing data compares with the performance of the RF surrogate trained using all design variables considered. This work is intended to inform design changes in future ICF experiments by augmenting the expert intuition and simulations results.
	\end{abstract}
	
	
	\section{Introduction}\label{Intro}
	
	The National Ignition Facility (NIF)\cite{moses2009ignition}, located at Lawrence Livermore National Laboratory, has the most energetic laser-based inertial confinement fusion (ICF) research device. Although there has been substantial progress in the last few decades, a self-sustaining burning plasma state has not been achieved~\cite{kline2019progress}. At NIF, the most often used ICF approach is called indirect drive. In the indirect drive fusion method, the lasers heat the inner walls of a gold cavity called a hohlraum containing the pellet, creating a superhot plasma that radiates x-rays. The x-rays rapidly heat the outer surface of the fuel pellet, causing a high-speed ablation of the surface material and imploding the fuel capsule in the same way as if it had been hit with the lasers directly. A symmetric compression of the capsule with radiation results in a hot, dense plasma where fusion reactions occur. The plasma ignites and the compressed fuel burns, releasing energy.
	
	Machine learning (ML) approaches have recently been applied to find insight into indirect drive ICF physics. Peterson et al. 2017~\cite{peterson2017zonal} use random forest regression to predict energy yield and to explore regions of high yield probability. They found that, contrary to expectations, ovoid implosions maximized yield instead of spherical ones. Humbird et al. 2018~\cite{humbird2018deep} and 2019~\cite{humbird2019transfer} use deep jointly-informed
	neural networks for training more predictive ICF models and propose a nonlinear technique for calibrating from simulations to experiments, or from low-fidelity simulations to high-fidelity simulations, via transfer learning, respectively. Hsu et al. 2020~\cite{hsu2020analysis} analyze relationships between implosion parameters and neutron yield informing ML models with physics. Nakhleh et al. 2020~\cite{nakhleh2020exploring} use random forest surrogate for prediction of neutron yield, neutron velocity, and other experimental outputs given a suite of design variables, along with an assessment of important relationships and uncertainties in the prediction model. 
	
\begin{figure*}[ht!]
		\centering
		\includegraphics[width=0.8\textwidth]{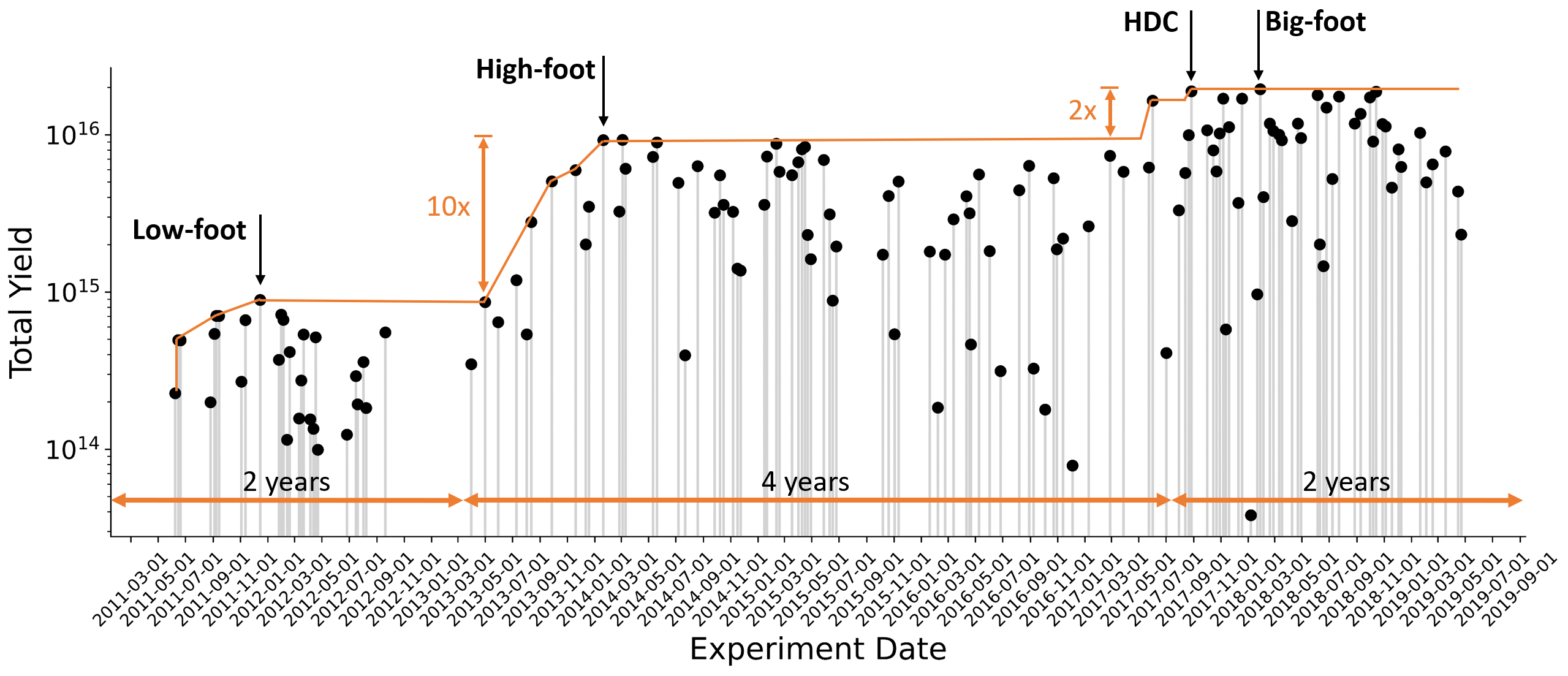} 
		\caption{Total yield obtained as a function of the experiment data. The experiments were conducted over a period of eight years. We can distinguish four important changes in design: low-foot, high-foot, high-density carbon (HDC), and big-foot. From 2011 to 2019 the highest yield obtained was increased 20 times.}
		\label{john_plot}
	\end{figure*}

	In this work, we use an experimental dataset of a total of 141 indirect drive ICF experiments from NIF. We consider 26 features associated with the ICF design variables and we study a single output, total yield, which is the neutron yield reported by the neutron detector, corrected based on its known measurement limitations. These experiments were conducted over a period of eight years, from 2011 to 2019. Figure~\ref{john_plot} shows the total yield obtained (logarithmic scale) as a function of the experiment date. During the first two years (31 experiments, low-foot) the reached yields were close to $10^{15}$ neutrons. During the following four years (61 experiments) the laser energy was increased and the number of pulse steps was reduced from four to three (high-foot). In the fourth year, the hohlraum gas fill density was reduced improving symmetry. This led to a roughly ten-time increase in yield ($10^{16}$ neutrons). From the sixth year until end of our study (49 experiments) the target material was changed to high-density carbon (HDC), and the fuel adiabat was increased along with the laser energy (big-foot). In this period the fill tube diameter was reduced from 10 $\mu m$ to 5 $\mu m$ mitigating the effect of 3D engineering features. During this period the yield achieved was doubled ($2 \times 10^{16}$ neutrons). 
	
	Table~\ref{tab4} summarizes the major changes in the NIF indirect drive ICF campaign from 2011 until 2019. Taking into account which physical variables were modified and to what degree during each design change can aid in understanding the outcomes of the ML techniques used in this work. As is clear from the description of the design stages, many of the changes in design variables were done simultaneously. In the context of ML, this leads to a correlation between design features. These relationships can obscure the assessment of the importance of particular design changes based solely on data-driven approaches. However, these correlations are manifestations of the physically-motivated underlying changes and by identifying natural groupings of features related to the same design changes, we can build quantitative features that allow the assessment of their importance. 
	
	\begin{table}
		\centering 
		\begin{tabular}{C{1.2cm} C{1.5cm} C{2.3cm} C{1.5cm}}
			Date & Design Change & Observed physics & Total Yield  \\
			\midrule\midrule
			6/2011 & Initial Design & Under-performance due to 3D effects  & $2 \times 10^{14}$ \\\\
			12/2011 & Low-foot & Total yield increases & $9 \times 10^{14}$ \\\\
			7/2013 & High-foot & Yield increases 10 times & $1.9 \times 10^{15}$ \\\\
			6/2015 & Hohlraum gas fill density reduction & Higher symmetry with improved consistency between the simulations and measurements & $9.0 \times 10^{15}$ \\\\
			7/2017 & HDC capsules along with improvement in surface finish  & Shorter laser pulse required, higher velocity implosions, reduction of instabilities. Yield is doubled & $1.8 \times 10^{16}$ \\\\
			10/2017 & Fill tube diameter reduction  & Mitigation of the effect of the 3D engineering features. Reduction in symmetry perturbation & Improved performance outside uncertainty bands \\\\
			3/2018 & Big-foot  & Yield is higher than from previous designs & $2 \times 10^{16}$ \\
			\midrule\midrule
		\end{tabular}
		\caption{Main changes to the NIF ICF design from June 2011 to April 2019~\cite{kline2019progress}.}\label{tab4}
	\end{table}
	
	The aim of this work is two-fold: i) predicting the output of interest based on the design variables, ii) identifying important design variables and meaningful design variable groupings to improve the prediction capability and untangle hidden design variables relationship to enhance the current understanding of the physics. 
	For this purpose an approach based on sparse component analysis (SPCA) is presented to identify compact groupings of design variables that represent underlying, physically-motivated changes. Principal component analysis (PCA) is used to determine the number of sparse components enforced when performing SPCA. This is done by choosing the minimum number of principal components that explain at least 95\% of the variance within the input dataset. Once the number of components is determined, SPCA is performed and then, using the obtained sparse components, the surrogate model is trained. 
	The performance of this approach is compared with the performance of the surrogate model trained directly on the physical variables instead of the physical components.
	
	The present manuscript is organized as follows. A detailed description of the design variables and the output of interest can be found in Section~\ref{exp}. In Section~\ref{ML} we describe in detail the SPCA-based approach considered including a brief analysis of the feature correlation matrix. In Section~\ref{results}, we present the results of the SPCA-based approach discussed. Finally, in Section~\ref{conclusions} we summarize this work and discuss future goals.
	
	\section{Experimental Data}\label{exp}
	
	The NIF dataset contains data from 141 ICF indirect drive experiments. In this work, 26 different design parameters are varied, such as laser energy, picket power, and hohlraum material. We focus our attention on predicting the total yield based on the remaining parameters. For clarity, we will give a brief outline of the ICF process and then discuss the relationship of ML dataset inputs to ICF physics.
	
	Indirect-drive ICF refers to the conversion of the laser energy to x-ray energy by a cylindrical hohlraum enclosing the capsule. The absorption of the x-rays at the capsule surface drives the capsule inward, compressing the fuel. As the fuel is compressed, it reaches extreme temperatures (5 $KeV$), densities (1000 $g/cm^3$), and pressures ($10^{12}$ $bar$), at which point fusion of the fuel occurs. Total yield is a measure of the energy generated through fusion during the ICF process.
	
\begin{figure}[ht!] 
		\centering
		\begin{subfigure}{1\columnwidth}
			\centering
			\includegraphics[width=\linewidth]{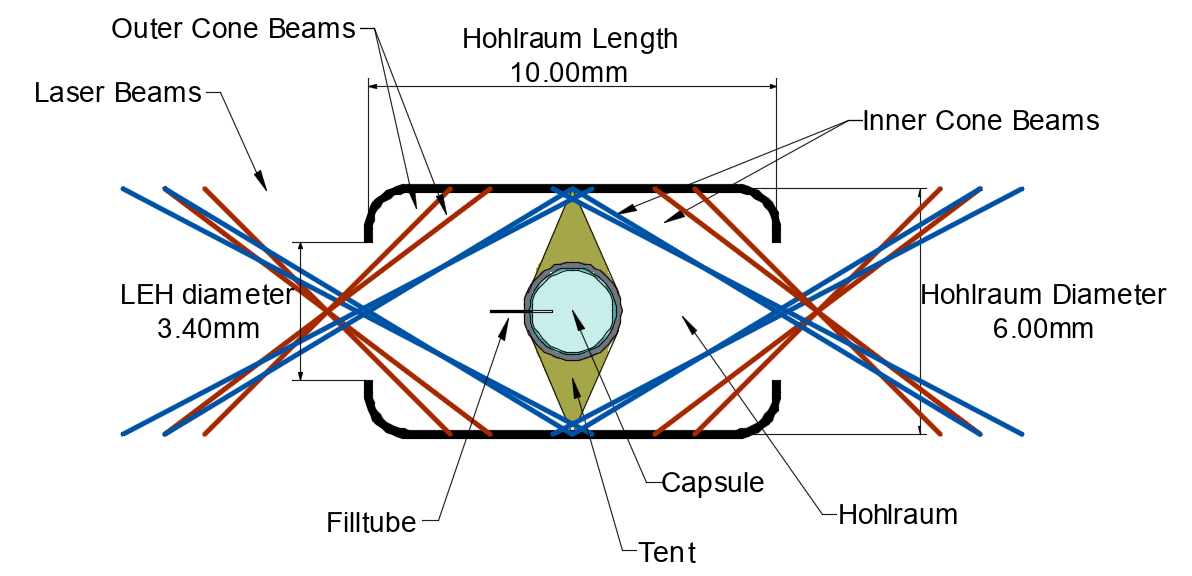}
			\caption{Schematic of the hohlraum.\label{hoh}}
		\end{subfigure} 
		\begin{subfigure}{0.8\columnwidth}
			\centering
			\includegraphics[width=\linewidth]{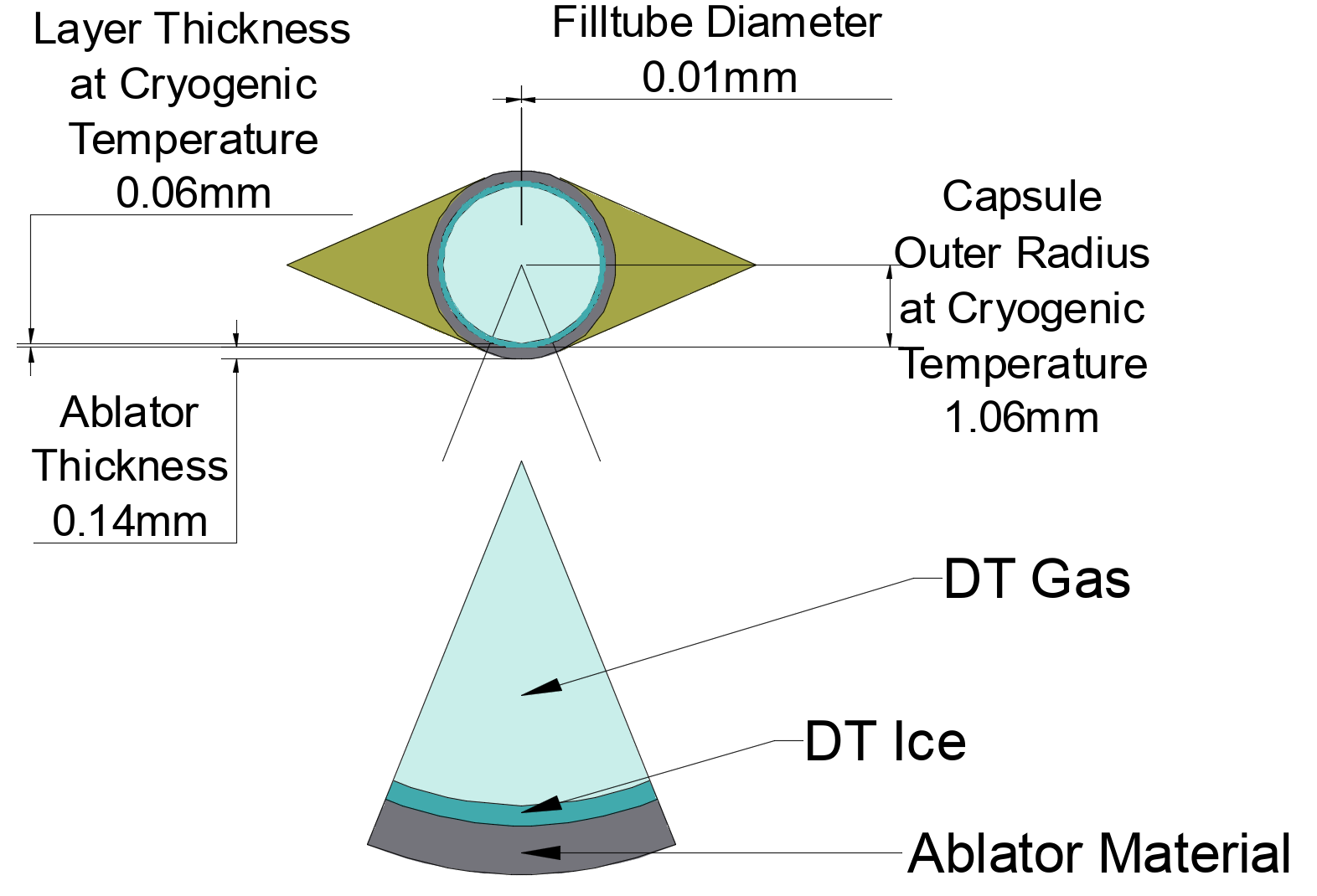}
			\caption{Schematic of the capsule.\label{cap}}
		\end{subfigure} 
		\caption{Schematic of the variables corresponding to the hohlraum and capsule physical group. The dimensions in $mm$ are taken as the mean dimension from the experimental dataset and they are included to give the reader an estimate.\label{phy_var}}
	\end{figure}

	The hohlraum is a cavity of a high atomic number, typically made of gold or uranium, whose walls are in radiative equilibrium with the radiant energy within the cavity. The hohlraum length and radius are approximately 10 $mm$ and 6 $mm$, respectively. In our experimental dataset, five input variables are directly related to the hohlraum: hohlraum material, hohlraum length, hohlraum diameter, density of the infill gas, and laser entrance hole (LEH) diameter. Figure~\ref{hoh} shows a schematic of the hohlraum. 
	
	The capsule is held by the hohlraum through a plastic film called tent. The capsule consists of an outer ablator shell encapsulating a deuterium-tritium (DT) mixture. The DT fuels the ignition process. The DT mixture takes the form of gas in the center of the capsule. Surrounding the gas, there is a layer of iced DT at cryogenic temperature. The capsule is the size of a pinhead ($2.4$ $mm$-diameter) and it contains around 150 $ \mu g$ of fuel. We consider six input variables that are directly related with the capsule: ablator thickness, ablator mass, tent thickness, layer thickness at cryogenic temperature, capsule outer radius at cryogenic temperature, and filltube diameter. Figure~\ref{cap} shows a schematic of the capsule. 
	
\begin{figure}[ht!]
		\centering
		\includegraphics[width=0.8\columnwidth]{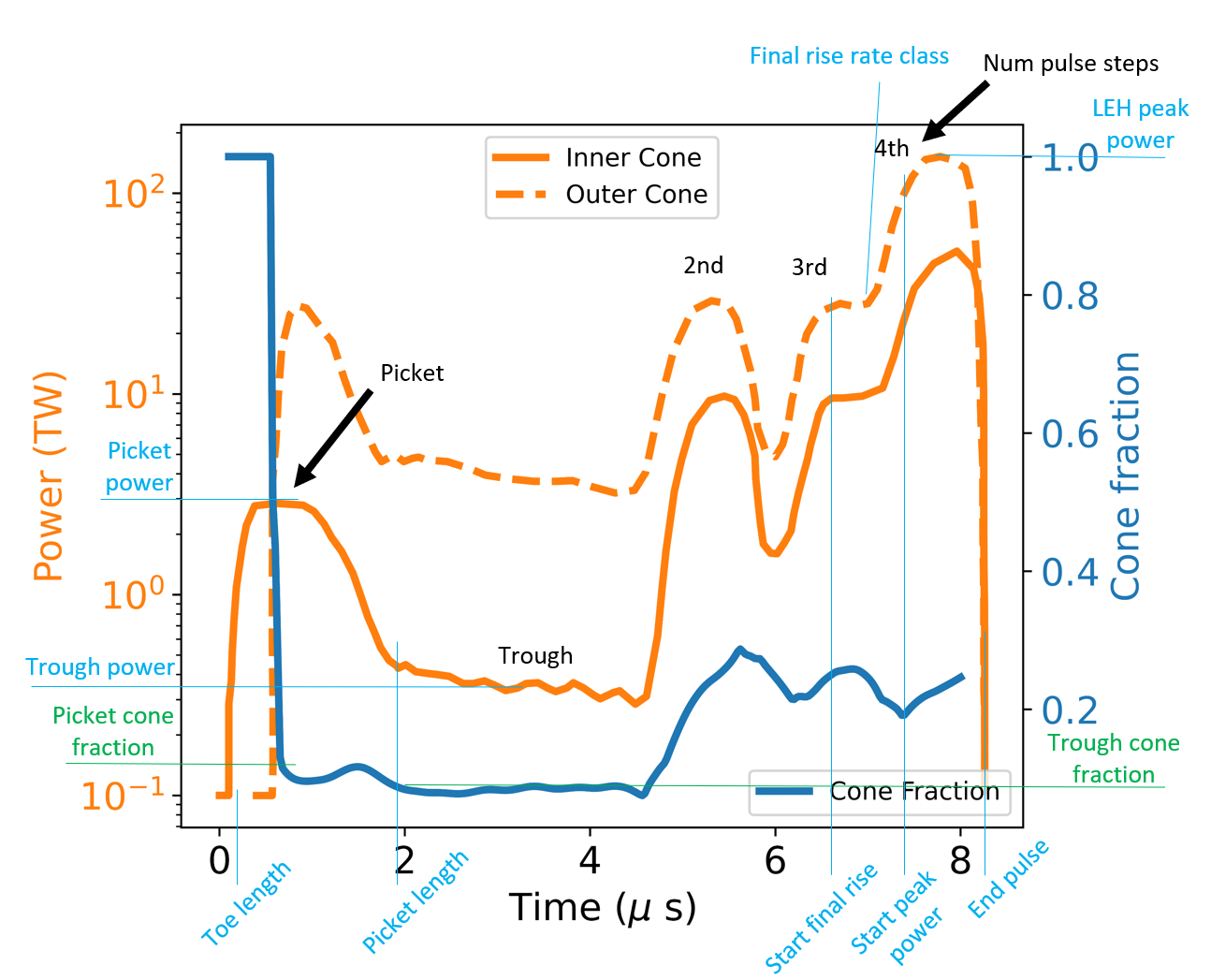} 
		\caption{Schematic of the laser power (inner and outer cone beams) and cone fraction as a function of time (laser physical group).}
		\label{laser}
	\end{figure}

	Finally, the remaining 15 input variables are directly related with the laser properties. These laser parameters mainly measure the time of events that occur between the start and the end of the laser pulse: picket length, toe length, start of the peak power, start of the final rise, and end of pulse. Some of them measure energy: LEH laser energy, LEH peak power, picket power, and trough power. Some measure cone fraction, which is the ratio of the inner cone beam power to the total pulse power: trough cone fraction and picket cone fraction. Others measure laser wavelengths: $\Delta \lambda_3 - \Delta \lambda_2$ and $\Delta \lambda_2$ which are used to control the cross beam energy transfer for the gas filled hohlraums~\cite{pickworth2020application}. Note that the number of laser-related inputs is three times higher than the number of hohlraum and capsule related inputs. Figure~\ref{laser} shows a schematic of the laser parameters. 
	
	In this work, we will refer to \textit{physical groups} as the three groupings of physical variables described above: the capsule physical group, the hohlraum physical group, and the laser physical group.
	
	\section{Machine Learning to Assess Relationships Between Design Features and Yield } \label{ML}
	
	\subsection{Correlations Between Design Features}\label{corr}
	
	The presence of strong correlations between design inputs can muddle the ability of purely data-driven methods to obtain clear interpretations of the strength of relationships between features and yield. These correlations are often driven by one underlying design change requiring the variation of a number of features. If information about feature correlations is not built into the surrogate model \textit{a priori}, highly-correlated features can be essentially equivalent in their use for prediction. This is important to remember with regard to using importance metrics as a tool to help inform a deeper, expertise-driven analysis, rather than interpreting the importance assessment causally.
	
\begin{figure}[ht!] 
		\centering
		\includegraphics[width=1\columnwidth]{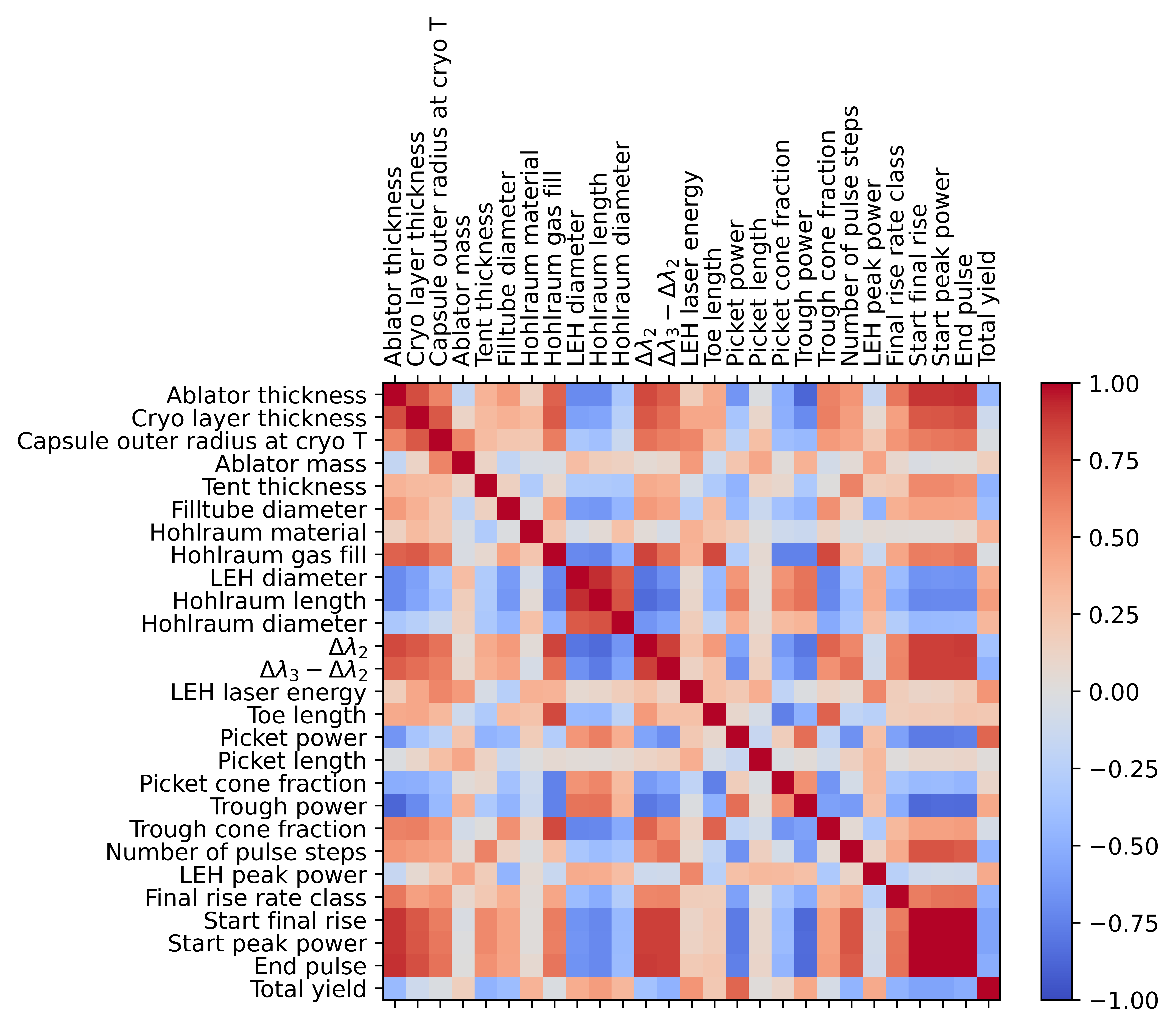} 
		\caption{Correlation matrix between the considered input variables and total yield. The colormap goes from -1 (perfectly negatively correlated) to 1 (perfectly positively correlated).}
		\label{correlations}
	\end{figure}

	As the first step to understanding feature relationships, we construct a matrix with correlations between variables to study the relationship between the 26 physical input variables and the output, total yield.  Figure~\ref{correlations} shows the correlations using a colormap that goes from -1 (perfectly negatively correlated) to 1 (perfectly positively correlated). The correlations were calculated using the Pearson correlation coefficient.
	%
	%
	In the figure, red represents positive correlations, white no correlation, and blue negative correlations.
	
	From the observation of Figure~\ref{correlations} we can highlight: 
	
	\begin{enumerate}
		\item Start final rise, start peak power, and end of pulse are highly positively correlated (more than 0.99). These pulse length related quantities are also highly negatively correlated with picket power (approximately -0.75) and trough power (approximately -0.85) because of the constraints of timing the shock.  
		\item Ablator thickness and $ \Delta \lambda_2$ are also highly correlated with start final rise, start peak power, and end of pulse (correlations of 0.89 and 0.87, respectively). As expected, ablator thickness presents a high negative correlation with picket power (-0.65) and trough power (-0.88). This negative correlation is associated with the high-foot design change, wherein the ablator thickness dropped roughly from $200 \mu m$ to $75 \mu m$ while the picket power increased from $15 TW$ to $40 TW$. 
		\item Hohlraum gas fill is highly positively correlated with toe length and $ \Delta \lambda_2$ with a correlation of 0.83 and 0.85, respectively. 
		\item LEH diameter has a high positive correlation with the hohlraum length (0.92 correlation). This has to do with a hohlraum design change during mid-2015, by the time that there was also a gas fill density reduction, that led to an increase in these two quantities.
		\item Ablator thickness and trough power are highly negatively correlated (-0.88 correlation). Trough power is highly negatively correlated with start final rise, start peak power, end of pulse (less than -0.88 correlation). $ \Delta \lambda_2$ is highly negatively correlated with hohlraum length with a correlation of -0.85. 
		\item We do not see extremely strong correlations with the output, total yield. However, the highest positive correlations with the output are with picket power (0.73 correlation) and LEH laser energy (0.52 correlation). The highest negative correlations are with start final rise (-0.56 correlation), start peak power (-0.56 correlation), and end of pulse (-0.51 correlation).
	\end{enumerate}
	
	\subsection{Feature Grouping with Sparse Principal Component Analysis}\label{SPCAnalisys}
	
	We can reduce the correlation between features in the set by finding data-driven groups of related features and building the surrogate model using features related to the latent groups. PCA (Pearson, 1901~\cite{pearson1901liii}) takes a set of observations of correlated variables and converts it into a set of linearly uncorrelated variables, called principal components, using an orthogonal transformation. The first principal component is the one that explains the largest variance within the data and each subsequent component has the highest variance providing it is orthogonal to the previous components.
	
	Consider the matrix $X$, where each of the $n$ rows represents a repetition of the problem and each of the $p$ columns represents a variable or feature. Let us assume also that $X$ mean is zero. We want to reduce the number of dimensions of the problem to $q$ ($p>q$) identifying the components that explain the maximum amount of data variance. These projections have to be orthogonal. To achieve this, we can solve two problems i) minimize the error (least-squares problem) or ii) maximize the variance (eigenvalue problem). In this work, we will focus on the second option, the maximum variance formulation~\cite{sharma1996applied}. PCA reconstructs the covariance matrix of $X$, $E(X^TX)$, by linear combinations of the original $p$ variables solving the following problem
	
	\begin{equation}\label{PCA}
	\begin{aligned}
	\max \quad & v_q^T(X^TX)v_q\\
	\textrm{s.t.} \quad & v_q^Tv_q=1\\
	&v_{q-1}^Tv_q=0,   \\
	\end{aligned}
	\end{equation}
	where $v_q$ denotes the $q$th eigenvector of $X^TX$ with loadings for each of the $p$ variables ($1 \leq q \leq p$). This problem can be solved using Lagrange multipliers and it leads to the following eigenvalue problem:
	
	\begin{equation}
	(X^TX-\lambda I)v_q=0.
	\end{equation}
	
	There are $p$ non-trivial eigenvalues $\lambda_q$ with $q=1, \dots, p$ and for each we get the corresponding eigenvector
	
	\begin{equation}
	v_q^T(X^TX)v_q= \lambda_q.
	\end{equation}
	
	The eigenvector $v_q$ is called the $q$th principal component and $\lambda_q$ is the variance explained by linear combination. When used for dimension reduction, PCA can be thought of as identifying a small set of latent modes driving the correlated variation in the features.
	
	A disadvantage of PCA is the fact that each principal component is a linear combination of all the original variables, i.e. all $p$ loadings $v_j$ with $j=1, \dots,p$ are different from zero. Thus it is often difficult to interpret the results if we are dealing with a high-dimensional problem.While multiple features can be changed at once to accommodate a particular physically-motivated goal in designing experiments, one would not expect these changes to always require a variation of \textit{all} the design parameters. SPCA provides an alternative to PCA with sparse loadings that lead to principal components that are a linear combination of a few physical variables balancing orthogonality and sparsity. By forcing some $v_j$ to be zero, the learned basis vectors will represent compact groups of input variables that have strong relationships between them. SPCA is used to emphasize variation and bring out strong patterns in the features for a dataset, often used to make data easy to explore and visualize. Hence, Eq.~\eqref{PCA} is modified as follows
	
	\begin{equation}\label{SPCA}
	\begin{aligned}
	\max \quad & v_q^T(X^TX)v_q\\
	\textrm{s.t.} \quad & v_q^Tv_q=1\\
	& \norm{ v_{q} } \leq k,   \\
	\end{aligned}
	\end{equation}
	with $k$ an integer such as $1 \leq k \leq p$. This optimization problem is not trivial and it requires a numerical optimizer. SPCA is used in this work to find meaningful groupings within the physical variables. SPCA compromises between orthogonality and sparsity. This leads to a loss of the total explained variance for the same number of principal components but a gain in sparsity. SPCA is a more expensive analysis because it involves an optimization process; however, in this case, the cost can be afforded. To compute the SPCA, in this work we use the decomposition function \textit{SparsePCA} from the Python library \textit{scikit-learn}~\cite{scikit-learn}, which has the variable projection approach as the optimization strategy. The numerical optimization requires more computational power than PCA; however it takes less than ten seconds for our problem. The sparsity controlling parameter is set to three and the amount of ridge shrinkage to apply in order to improve conditioning is set to $10^{-2}$. 
	
	PCA and SPCA are designed to find sparse components through a linear transformation. However, if the relationships between variables are not linear, these methods will not be able to capture the relationships correctly. Our dataset presents highly linear relationships and there are no complex nonlinear relationships between variables. This claim is supported by the fact that the prediction performance is not deteriorated by using SPCA components to train the RF surrogate as shown in Section~\ref{results}.
	
	\section{Approach}\label{approach}
	
	Our approach has three basic steps: First, PCA is used to determine the number $n$ of sparse components enforced when performing SPCA. This is done by choosing the minimum number of principal components that explain at least 95\% of the variance within the input dataset. Because SPCA balances sparsity and orthogonality there is a small reduction in the explained variance when moving from PCA to SPCA; however this reduction is small~\cite{zou2006sparse}. Second, once the number of components is determined, the SPCA is performed. Third, using the obtained sparse components, a surrogate model is trained. 
	In this work we have chosen RF regression as a surrogate model because it outperforms neural networks and Gaussian process predicting on the test data
	
	\tikzstyle{block} = [rectangle, draw, 
	text width=20em, text centered, rounded corners, minimum height=4em]
	\tikzstyle{line} = [draw, -latex']
	
	
	\section{Results}\label{results}
	
	\subsection{Preliminary Principal Component Analysis} \label{sec_PCA}
	
	In order to determine the number of sparse components to be imposed when calculating SPCA, we rely on PCA. We choose the number of sparse components to be the one that explains at least 95\% of the variance in the input features according to PCA. To perform PCA, we use the Python scikit-learn PCA decomposition package. The cumulative explained variance (CEV) is calculated using Eq.~\eqref{ExV},
	
	\begin{equation}\label{ExV}
	\textnormal{CEV} (c) =
	\begin{cases}
	0 & \text{if $c = 0$ } \\
	\sum_{j=1}^c  \frac{\lambda_j}{\sum_{i=1}^n \lambda_i} & \text{if $c \geq 1$},
	\end{cases}
	\end{equation}
	where $c$ is the principal component number that we are considering, $n$ is the total number of principal components (in this case we set $n=26$ which is the maximum number of principal components possible in a dataset with 26 features), $\lambda_i$ is the eigenvalue corresponding to the principal component number $i$ with $i \in \left\lbrace 1,2, \dots, n\right\rbrace $,  and $\lambda_j$ is the eigenvalue corresponding to the principal component number $j$ with $j \in \left\lbrace 1,2, \dots, c\right\rbrace $. We selected the number of components $c$ such that CEV$(c)=0.95$. In this case, PCA run on 100 different randomly selected training/test datasets determines that $10.5 \pm 0.5$ components should be used to explain at least 95\% of the dataset variance. Based on these results, SPCA is set to build 11 sparse components (i.e. $c=11$).
	
	
	\subsection{PCA-based SPCA Components}
	
	\begin{figure*}[ht!] 
		\centering
		\includegraphics[width=0.7\textwidth]{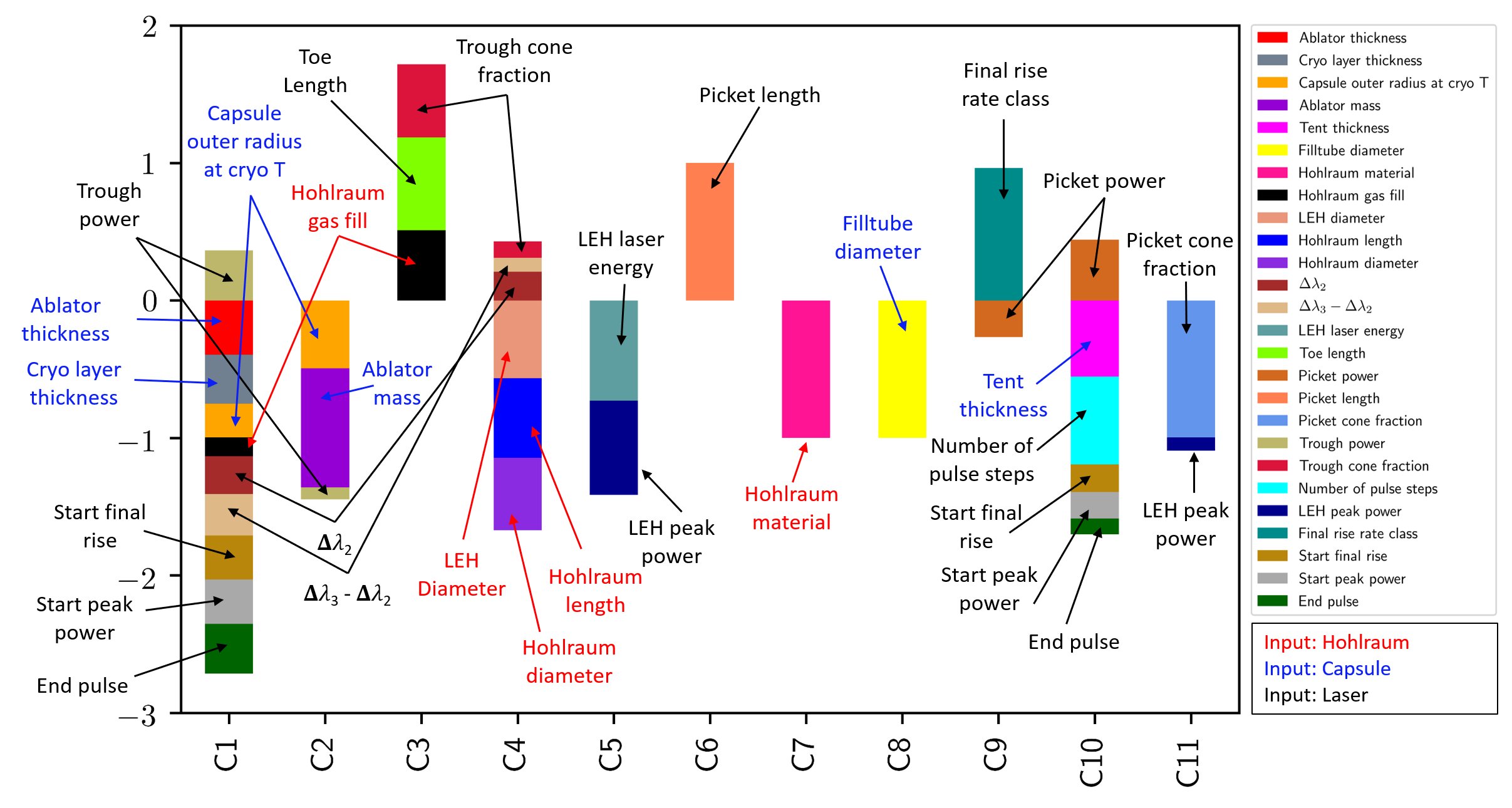} 
		\caption{The 11 components determined by SPCA. The horizontal axis corresponds to the sparse component name, while the vertical axis measures the weight that each physical variable has in a particular sparse component. Each color represents a physical variable (26 in total). The red labels represent physical variables from the hohlraum physical group, the blue labels from the capsule physical group, and the black labels with the laser physical group.}
		\label{CompSPCA}
	\end{figure*}
	
	Based on the analysis of Section~\ref{sec_PCA}, 11 components were imposed while computing SPCA using the entire dataset. The 11 resulting components are shown side-by-side in Figure~\ref{CompSPCA}. In the figure, each color represents a physical variable (26 in total) which is listed on the upper right legend. The horizontal axis corresponds to the sparse component name. On the vertical axis, the width measures the weights of each physical component. The red labels represent physical variables from the hohlraum physical group, the blue labels from the capsule physical group, and the black labels from the laser physical group. The weights in a particular component are scaled such the total length of the vector of those weights is equal to one (Euclidean length). Note that components C6, C7, and C8 are all of length one, C5 length is approximately 1.4 ($\sqrt{2}$), and C1 length is approximately 3.2 ($\sqrt{10}$). Hence within a component, the weights give the relative importance of a feature to that component but comparison between weights in different components is not meaningful.
	
	SPCA finds three single components and eight combined components (11 total). Of the three single components, C6 is composed of a physical variable coming from the laser physical group (picket length), C7 from the hohlraum physical group (hohlraum material), and C8 from the capsule physical group (filltube diameter). From the eight combined components, two are composed mainly of a single physical variable coming from the laser physical group C9 (final rise rate class) and C11 (picket cone fraction). We also note that these two combined components are entirely composed of physical variables coming from the laser physical group. From the remaining six components, five of them are mainly composed of a few physical variables which are C2, C3, C4, C5, and C10. C2 is mainly composed of ablator mass and capsule outer radius at cryogenic temperature, both variables from the capsule physical group. C3 is composed of $2/3$ of variables from the laser physical group (trough cone fraction and toe length) and $1/3$ from the hohlraum physical group (hohlraum gas fill). C4 is mainly composed of physical variables related to the hohlraum physical group (LEH diameter, hohlraum diameter, and hohlraum length). C5 is composed of laser-related physical variables (LEH laser energy and LEH peak power). Finally, C10 is roughly $2/3$ composed of variables from the laser physical group (picket power and number of pulse steps) and $1/3$ from the capsule physical group (tent thickness). The component that is left, C1, has many features with a similar weight each. It has approximately $1/3$ of the weight that comes from capsule related physical components (ablator thickness, layer thickness at cryogenic temperature, and capsule outer radius at cryogenic temperature) while the rest comes mainly from laser associated quantities (trough power, $\Delta \lambda_3 - \Delta \lambda_2$, $\Delta \lambda_2$, start final rise, start peak power, end pulse) except for hohlraum gas fill which is a hohlraum related physical component. Nine variables are present in more than one sparse component, but in all cases, their presence is only substantial in one component.
	
	\subsection{Importance Metrics} \label{importance}
	
	Understanding the relationships between controllable design variables and experimental output(s) of interest, as well as the relationships between the newly found groupings and the output, can be useful for gaining new physical insights. We measure importance through a metric called accumulated local effects (ALE)~\cite{apley2016visualizing}. ALE describes how features or groups of features influence the prediction of an ML model on average analogous to the main effects in variance-based function decomposition\cite{saltelli_global_2008}. ALE averages the changes in the predictions and accumulates them over the grid, providing unbiased estimates of the main effect even when variables are correlated. The relative importance of features can be compared through the functional variance of their main effects. 
	
\begin{figure}[!ht] 
		\begin{subfigure}{0.9\columnwidth}
			\centering
			\includegraphics[width=\linewidth]{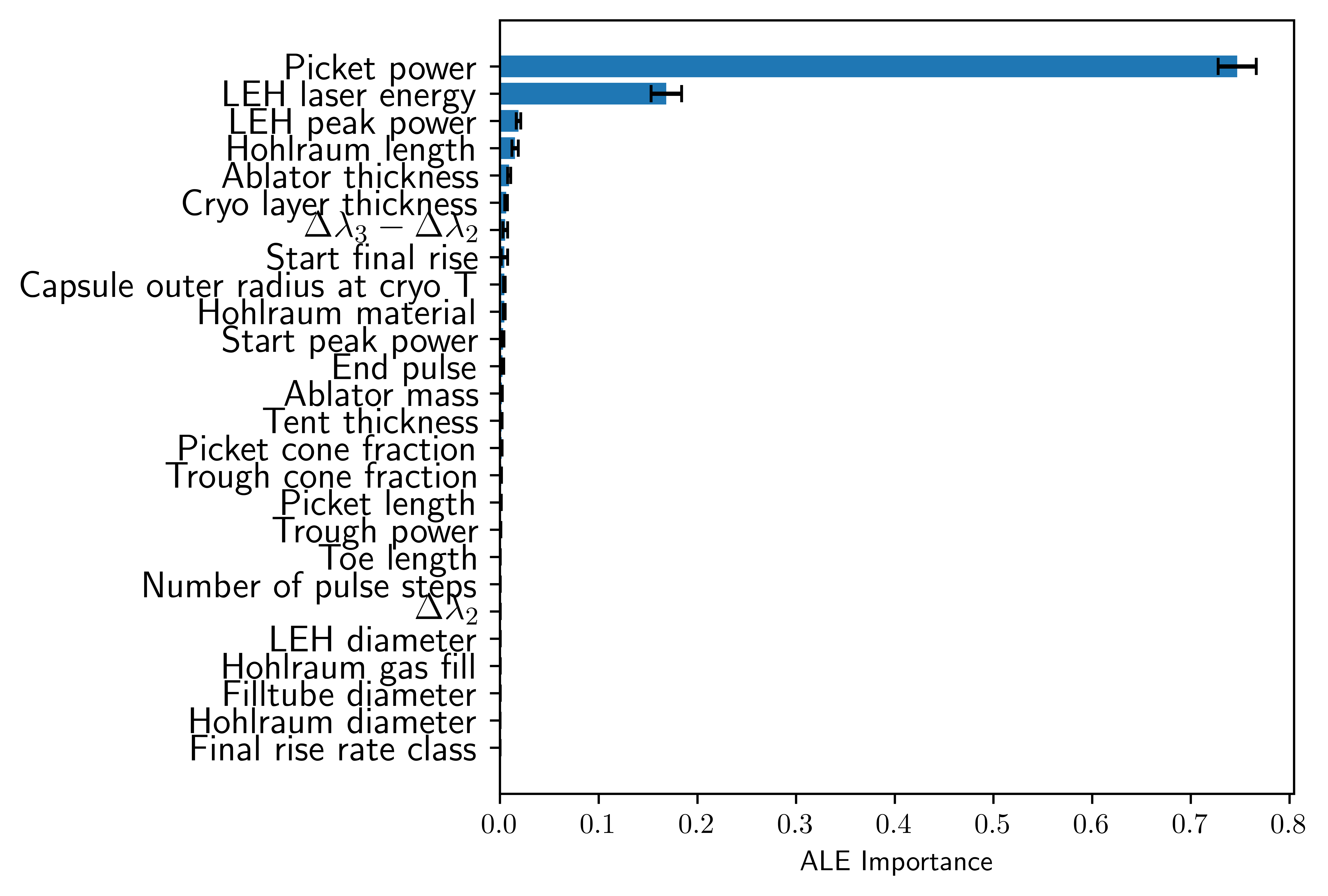}
			\caption{ALE importance results for the RF surrogate model trained using the 26 physical design inputs (RF26). The error bars correspond to two standard deviations resulting from 100 randomly selected training/test datasets. \label{ALE_one_column}}
		\end{subfigure} \\
		\begin{subfigure}{0.9\columnwidth}
			\centering
			\includegraphics[width=\linewidth]{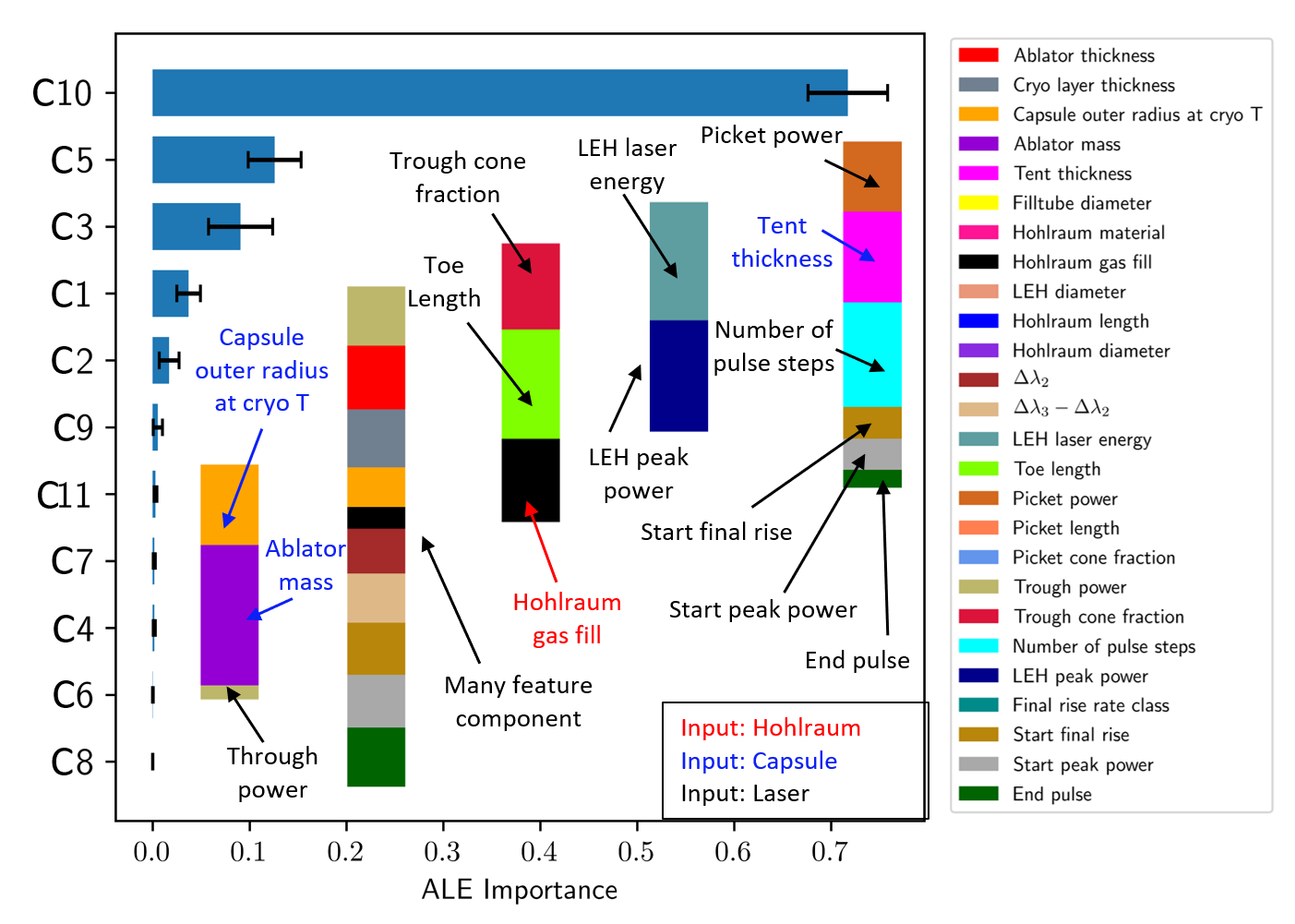}
			\caption{ALE importance results for the RF surrogate model trained using the 11 sparse principal components (C) found by SPCA. The component number corresponds with the one used in Figure~\ref{CompSPCA}. The error bars correspond to two standard deviations resulting from 100 RF fitting iterations.\label{ALE_SPCA}}
		\end{subfigure} 
		\caption{ALE importance metric. \label{fig:ALE26}}
	\end{figure}

	Figure~\ref{ALE_one_column} shows ALE importance results from an RF surrogate trained for total yield prediction using 100 different randomly selected training datasets (80 \% of the full dataset). As the figure shows, picket power is by far the most important feature for predicting total yield if the RF surrogate is trained using the 26 physical design inputs. This may be related to two physics-based reasons. On the one hand, the design change from low-foot to high-foot increased the picket power to improve implosion stability. This higher picket power puts the capsule on a higher adiabat, making it a bit harder to compress but also reducing some of the susceptibility to instabilities. On the other hand, after the design change from high-foot to HDC designs the picket power also increased leading the first shock to be above the melt conditions and resulting in a reduction of voids and imperfections that are common in these capsules. In both these cases, the picket power itself was not the only change, but part of a larger design variation. Picket power is followed in importance by LEH laser energy, LEH peak power and hohlraum length. This is consistent with what we found through the correlation matrix (Figure~\ref{correlations}), where picket power and LEH laser energy are highly positively correlated with total yield.
	
\begin{figure}[ht!] 
		\centering
		\includegraphics[width=0.8\columnwidth]{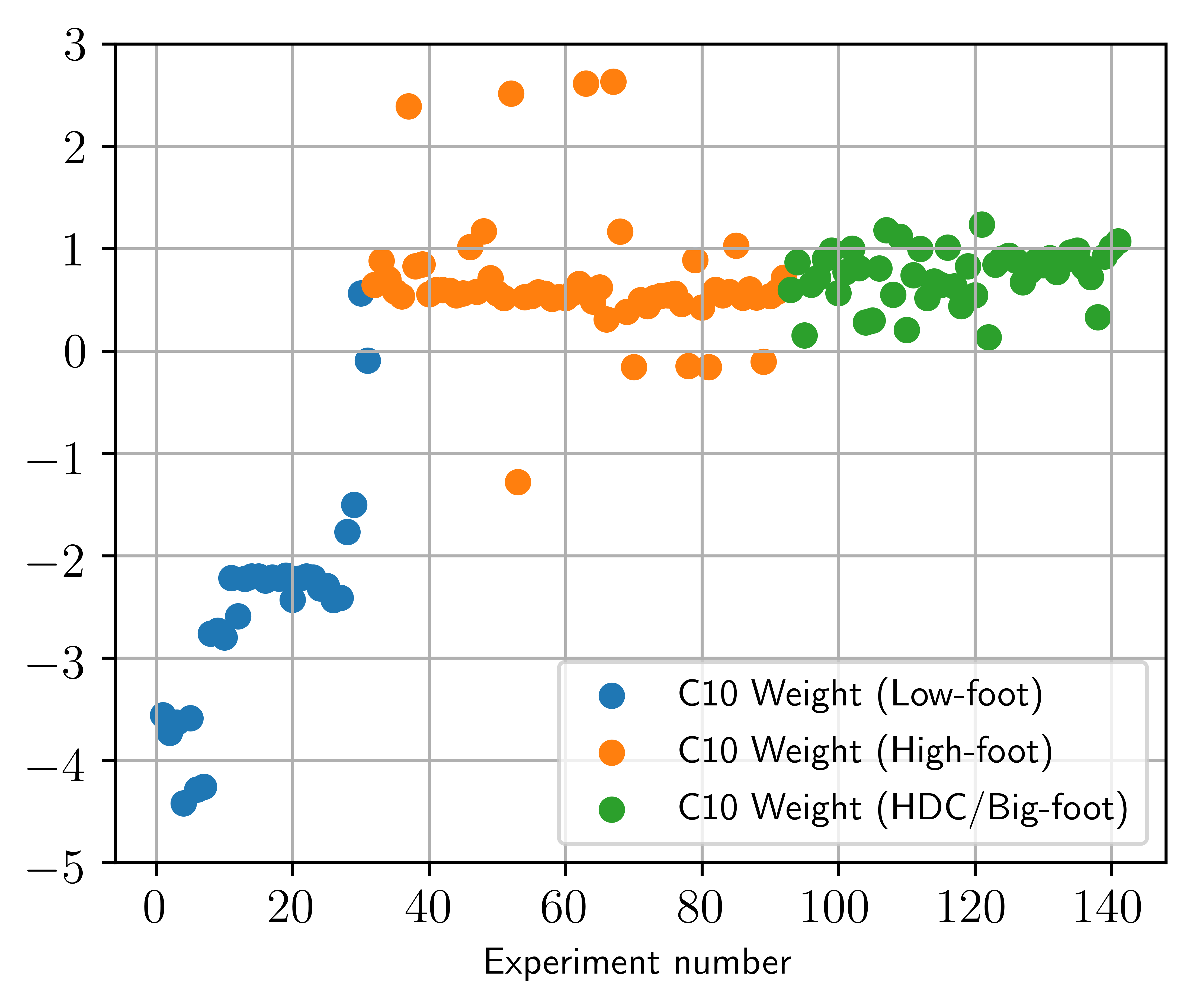} 
		\caption{Weight of the most important sparse principal component, C10, for each of the 141 experiments. The coloring differentiates the most relevant design changes (Low-foot, High-foot and HDC/Big-foot). C10 appears to be indicative of the change from low-foot to later designs. }
		\label{weights}
	\end{figure}

	Figure~\ref{ALE_SPCA} shows the ALE importance metric calculated using the RF surrogate model trained using the 11 sparse principal components. The component number corresponds with the one used in Figure~\ref{CompSPCA}, however, a reminder of the composition of the five most important sparse components is included to aid in visualization. ALE finds that C10 is by far the most important component. This component is mainly composed of design features related to the laser pulse shape and timing, with picket power as one of the three most predominant physical variables, as well as number of pulse steps, start of the final rise, start of peak power, and end of pulse. It is also curiously involved tent thickness. This component appears to capture the variation between the early low-foot and later designs. Figure~\ref{weights} shows the weights of C10 for the 141 experiments considered in this work. The coloring differentiates the most relevant design changes (Low-foot, High-foot and HDC/Big-foot). The weights for C10 SPCA largely separate the low-foot from other designs. The identification of its importance for predictive capability reflects the value of the change away from low-foot designs, which was identified as one of the most important design changes by subject-matter experts. The inclusion of tent thickness results from its reduction from $300$ $nm$ to $50$ $nm$ roughly at the same time of this change. 
	
	The second most important component is C5, composed of LEH laser energy and LEH peak power, which are both highly correlated with total yield (0.52 and 0.40, respectively). LEH laser energy has the second-highest positive correlation with total yield after picket power, which has a positive correlation of 0.73. This indicates that, while variations in timing were the most predictive of performance, the changes in the strength of the drive were also of importance. C3, C1, and C2 follow in importance. C3 is mainly composed of laser associated physical variables and C2 with capsule associated physical variables whereas C1 is a many feature component. These three components together have roughly 10\% of the importance. 
	
	Figure~\ref{CompSPCA} and therefore Figure~\ref{ALE_SPCA} are obtained by training SPCA on the entire dataset. That is, there is not an 80\% training - 20\% test split when conducting SPCA in this case. If the training data is varied we noticed the following: 1) most of the components will remain without noticeable changes in composition (eight sparse components or more), 2) the changes in the component composition are smooth and in most cases, they correspond with a few variables passing from one component to another, and 3) the most important components are composed of the most important physical variables, that is, picket power, LEH laser energy, LEH peak power, and hohlraum length, among others. 
	
	\subsection{Performance}\label{performance}    
	
	The RF surrogate is trained using the original dataset and the sparse dataset. To measure goodness of fit, we use relative RMSE, calculated as:
	
	\begin{equation}\label{RelRMSE}
	\textnormal{Relative RMSE} =\sqrt{\frac{1}{N} \sum_{i=1}^{N} \left( \frac{\hat{y}_{i} - y_{i}}{y_{i}} \right)^2} \times 100,
	\end{equation}
	where $\hat{y}_{i} $ is the prediction at the $i$th data point, $y_{i}$ is the training/test data corresponding to the $i$th data point. $N$ is the total number of data points considered. Note that we report the relative RMSE as a percentage quantity. Table~\ref{tab1} reports the relative RMSE computed for the training/test data set as the average of 100 different training/test datasets to account for stochastic variation in different RF iterations. The performance on the training data is consistent and equal to 0.8\% relative error. The performance on the test data is $\approx$ 2\% for both the original RF trained with the 26 physical design variables (RF26) and the RF trained using the 11 components found by SPCA. The table shows an overall excellent RF surrogate performance but also shows that SPCA gives 11 sparse components that can be used to achieve the same accuracy obtained using the original 26 variables.
	
	\begin{table}[h!]
		\centering
		\begin{tabular}{l|c|c}		
			& \multicolumn{2}{c}{	Mean Rel. RMSE $\pm 2 \sigma$ [\%]}\\
			\hline
			& Training data & Test data\\
			\hline
			\hline
			RF26   & 0.8 $\pm$ 0.1  &  2.0 $\pm$ 0.1\\
			RF-SPCA & 0.8 $\pm$ 0.1   & 1.8 $\pm$ 0.1 \\
			\hline
		\end{tabular}
		\caption{Mean relative RMSE and two standards deviations ($\sigma$) associated with 100 different randomly selected training/test datasets for the two surrogates considered (RF26 and RF-SPCA). The mean relative RMSE is reported for the training data and the test data. The performances are very similar.}
		\label{tab1}
	\end{table}
	
	\section{Conclusion} \label{conclusions}
	In this work, we use machine learning (ML) methods for variable grouping and importance analysis to identify complex physical relationships to augment expert intuition and simulation results for the optimal design of future ICF experiments. We build sparse components through sparse principal component analysis (SPCA). SPCA tends to find groupings that are related to the physical origin of the variables (laser, hohlraum, and capsule), as well as components associated with design changes that dramatically influenced the performance. The accumulated local effects (ALE) importance results extracted from a random forest (RF) surrogate trained with the original 26 variables demonstrate two important variables: picket power and LEH laser energy. Both variables are highly positively correlated with the studied output, total yield. ALE importance results from an RF surrogate trained with the 11 sparse components give more balanced importance between components, highlighting groups that are highly correlated with the total yield for physical reasons, like LEH laser energy, as well as components associated with design changes that dramatically influenced the performance, like number of pulse steps. We find that by considering just 11 sparse principal components, we can predict total yield using an RF surrogate as accurately as with the original 26 variables (within 2\% error) for both training and test datasets. The overall performance is outstanding, achieving a relative error of less than 1\% on training data and less than 2\% on test data across all approaches.  
	
	Although we have made progress identifying design input groupings, we will continue studying ML-based clustering methods and discussing the findings of such analyses with ICF experts. Future work will extend this analysis to other experimental outputs of interest, such as implosion velocity and areal density from down-scattered ratio, and investigate the connection between simulations and experiments in order to create tools that can provide uncertainty estimates where experimental data is not available.
	
	\section*{Acknowledgments}
	
	This work was supported by the U.S. Department of Energy through the Los Alamos National Laboratory. Los Alamos National Laboratory is operated by Triad National Security, LLC, for the National Nuclear Security Administration of U.S. Department of Energy (Contract No. 89233218CNA000001). Approved for public release LA-UR-20-28715. We would like to thank Dr. Otto Landen for the use of his NIF experimental database as the source for this work. We would also like to thank Grant Meadors and Josh Sauppe for their technical feedback.
	
	\bibliographystyle{IEEEtran}
	\bibliography{IEEEabrv,bibliography}
	
\end{document}